\documentclass[10pt,twocolumn]{revtex4}
\usepackage{amsmath}
\usepackage{latexsym,amssymb,epsfig,graphicx, bm}

\begin{document}

\title{Ferroelectric ordering in chiral smectic C$^*$ liquid crystals \\determined by nonchiral
intermolecular interactions}

\author{M.A. Osipov$^1$ and M.V. Gorkunov$^{1,2}$}
\affiliation{$^1$Department of Mathematics, University of
Strathclyde, Glasgow G1 1XH, UK \\ $^2$Institute of Crystallography,
Russian Academy of Sciences, 119333 Moscow, Russia}

\begin{abstract}
General microscopic mechanism of ferroelectric ordering in chiral
smectic C* liquid crystals is considered. It is shown that if the
mesogenic molecules have a sufficiently low symmetry, the
spontaneous polarization is proportional to one of the biaxial
vector order parameters of the smectic C phase. This order parameter
may be determined by intermolecular interactions which are not
sensitive to molecular chirality. At the same time, the polarization
is also proportional to a pseudoscalar parameter which vanishes if
the molecules are nonchiral. The general statistical theory of
ferroelectric ordering is illustrated by two particular models. The
first model is based on electrostatic quadrupole-quadrupole
interactions, and  it enables one to obtain explicit analytical
expressions for the spontaneous polarization. In the second model,
the molecular chirality and polarity are determined by a pair of
off-center nonparallel dipoles. For this case, the spontaneous
polarization is calculated numerically as a function of temperature.
The theory provides a more general interpretation of the previous
approaches including the classical Boulder model.
\end{abstract}

\maketitle

\section{Introduction}

Ferroelectric smectic liquid crystals are unique systems where the
spontaneous polarization is determined by molecular chirality.
Ferroelectric ordering in the chiral smectic C$^*$ phase has been
predicted theoretically about three decades ago\ \cite{Meyer1}, and
both ferro- and antiferroelectric phases continue to attract a
significant attention because of their unusual structure and
properties, and because of their applications in electro-optical
devices \cite{FrankJan}. Ferroelectricity is observed only in tilted
smectic phases. In these systems the spontaneous polarization is
induced by the tilt and does not appear self-consistently like in
conventional solid ferroelectric materials. In every tilted layer of
a chiral smectic phase, the polarization appears in the direction of
the polar $C_2$ symmetry axis which is perpendicular to the tilt
plane. In the bulk tilted phase molecular chirality also results in
the formation of the macroscopic helical structure. In this
structure, the direction of the tilt rotates while moving along the
z-axis which is perpendicular to the smectic layers. As a result,
chiral tiled smectics are characterized by the helical distribution
of the spontaneous polarizations and thus may also be called
'helielectric'. One notes that recently fero- and antiferroelectric
ordering has been found in a novel class of smectic liquid crystal
phases formed by achiral bent-core molecules. In those phases the
spontaneous polarization is not induced by the tilt and thus appears
also in orthogonal smectic phases \cite{bananareview}.

In the synclinic smectic C$^*$ phase, the direction of the tilt in
adjacent layers is practically the same and thus the spontaneous
polarization only slowly varies from layer to layer. In contrast, in
the anticlinic smectic C$^*_A$ phase the direction of the tilt
alternates from layer to layer together with the polarization
creating the structure with an antiferroelectric type ordering. In
addition, many chiral smectic materials exhibit a sequence of the
so-called intermediate phases in a narrow temperature interval
between the synclinic ferroelectric Smectic C$^*$ and the anticlinic
antiferroelectric Smectic C$^*_A$ phase. Intermediate smectic phases
are characterized by a 3D chiral distribution of the spontaneous
polarization within the unit cell of 3 or 4 smectic layers
\cite{FrankJan}. Recently, the remarkably wide intermediate phases
have also been discovered in mixtures of synclinic and anticlinic
smectics where they can exist in a broad temperature range of up to
$50^0$ \cite{Helen,Jan}. Finally, ferroelectric ordering also exists
in smectic I and F phases which are characterized by some in-plane
positional or hexatic order.

In spite of all diversity of tilted smectic phases with ferro-,
antiferro- and ferrielectric ordering, the underlying mechanism is
always related to the induction of the polarization by the tilt in
individual chiral smectic layers. Complex structures with a
polarization distribution along the direction perpendicular to the
layers appear due to interactions between the molecules in different
layers (see, for example, \cite{cepic,emelos}) which are generally
weaker than intermolecular interactions within the same layer.

The detailed microscopic mechanism of ferroelectric ordering in
tilted smectics which is responsible for the induction of the
polarization by the collective molecular tilt has been the issue of
debate during the past two decades. In particular, the role of
molecular chirality has not been completely clarified. On the one
hand, there is a general agreement that the spontaneous polarization
in tilted smectics cannot exist without molecular chirality, i.e. at
least a part of molecules must be chiral. On the other hand the role
of chiral intermolecular interactions remains unclear. Some
molecular models of the ferroelectric smectic C$^*$ phase are based
on the assumption that the spontaneous polarization is directly
determined by appropriate interactions between chiral molecules
including, for example, interactions with a molecular chiral center
\cite{osipov84,ospikrev,MCLCreview,ossteg,zeks}. These models have
been developed using an analogy with cholesteric liquid crystals
where the helical twisting power is determined by chiral (albeit
nonpolar) intermolecular interactions
\cite{Vertogen,osemelchol,molreview}. Other models are based on a
different microscopic mechanism
\cite{goosens1,goosens2,boulder1,boulder2,photinos1,photinos2,giesselmann}which
also requires molecular chirality, but, at the same time, takes into
consideration only nonchiral intermolecular interactions. Some of
these models are rather qualitative, but the underlying microscopic
mechanism is essentially the same.

\begin{figure}
\centering\includegraphics[width=8cm]{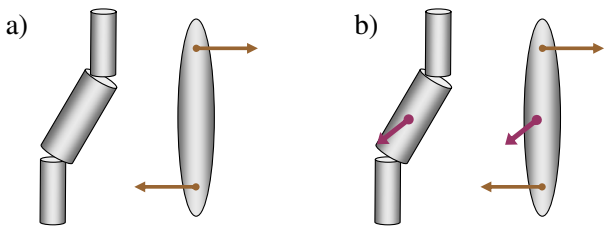} \caption{(Color
online) a) Simple models for nonchiral molecules of the $C_{2h}$
symmetry used by Wulf \cite{wulf} and Goosens
\cite{goosens1,goosens2}. b) Corresponding models for chiral
molecules where chirality is determined by the additional dipole
perpendicular to the molecular plane. }\label{Fig1}
\end{figure}

The first extended description of this mechanism has been given by
Goosens \cite{goosens1,goosens2} who considered the electrostatic
interaction between model molecular quadrupoles composed of two
antiparallel dipoles  which are perpendicular to the long molecular
axis ( see Fig.~\ref{Fig1}).
Goosens has shown that the polarization in the smectic C$^*$ phase
may be induced by the nonchiral quadrupole-quadrupole electrostatic
interaction provided the molecules possess the additional dipole in
the direction perpendicular to the molecular plane. The latter
dipole is responsible for the molecular chirality in this simple
model, and the spontaneous polarization is proportional to the
magnitude of the dipole and to the novel order parameter of the
smectic C \cite{goosens1} phase which is related to the relatively
low symmetry of the molecule presented in Fig.~\ref{Fig1}. One
notes, however, that the papers of Goosens are focused into one
particular model, and do not contain any general theory of
ferroelectric ordering or general expressions for the spontaneous
polarization.

\begin{figure}
\centering\includegraphics[height=4cm]{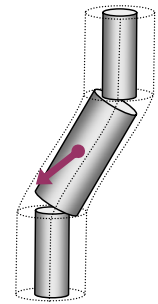} \caption{(Color
online) Schematic of biaxial molecule in the curved binding site as
assumed in the Boulder model}\label{Fig2}
\end{figure}

A more general and a very successful model has been proposed by the
Boulder group \cite{boulder1,boulder2}. In the Boulder model the
molecules of the zig-zag shape (see Fig.~\ref{Fig2}) are ordering in
the so-called binding cites which have the same point symmetry as
the smectic C phase itself. Then transverse molecular dipoles are
ordered in the particular direction perpendicular to the tilt plane
simply because the zig-zag molecule fits into the binding cite of
the same shape only for a particular direction of the transverse
dipole. The Boulder model has been successfully used to describe and
predict the value and sign of the spontaneous polarization for a
significant number of chiral smectic materials. This indicates that
the corresponding mechanism of the ferroelectric ordering may be
predominant at least for conventional smectics C$^*$. One notes also
that the symmetry of a zig-zag molecule is exactly the same as that
of the molecule with two equal antiparallel dipoles considered by
Goosens. The interaction between the molecule and the binding cite ,
which is responsible for the polar order, is also nonchiral in
nature because the binding cite itself is nonchiral. Boulder model
emphasizes the steric mechanisms of the ordering, but the idea
behind the model is much more universal. It is shown in this paper
that in the context of a rather general molecular-statistical theory
the concept of the 'binding cite' corresponds to the average
one-particle mean-field potential which is created by all other
molecules of the medium, and which reflects the symmetry of the
smectic C phase.

Terzis, Photinos, Samulski et.al. have developed a similar model
\cite{photinos1,photinos2}. This detailed model is based on a
mean-field -like one-particle orientational potential for each
molecular segment and, similar to the Boulder model, involves a
summation over the molecular conformations. Using this model
Terzis et.al. have obtained good quantitative results for the
spontaneous polarization of a number of ferroelectric smectics
C$^*$ \cite{photinos2}.

 Existing microscopic models for ferroelectric smectics C$^*$ have
 played an important role in the development of the theory of such
 materials. At the same time, from the point of view of theoretical
 physics, these models are too detailed in terms of a
 molecular structure and a particular choice of a coupling with the macroscopic
  environment. As a result, the models only indirectly address
 some of the more general physical problems related to the origin of
 ferroelectricity in tilted smectic phases including, for
 example, the description of the symmetry properties of relevant
 model interaction potentials and an interplay between the
 spontaneous polarization and the order parameters of the smectic C$^*$
 phase.

 In this paper we develop a general mean-field
 molecular-statistical theory of ferroelectric ordering in the the
 smectic C$^*$ phase based on the general mechanism described above,
 i.e.,  we consider the ferroelectric ordering in chiral smectics
 caused by {\it nonchiral} intermolecular interactions. We obtain a
 simple expression for the model interaction potential which may be
 responsible for ferroelectric ordering in tilted smectics, and
 interpret the general microscopic mechanism of the ordering
 mathematically using the concept of the pseudovector order
 parameter. The results of the theory will be used to obtain
 explicit expressions for the spontaneous polarization in the
 smectic  C$^*$ phase composed of biaxial molecules with
 quadrupole-quadrupole interaction potential. The spontaneous
 polarization together with other order parameters of the smectic C$^*$
 phase will also be calculated numerically for another two
 interaction model; Finally, we will consider in detail the
 molecular origin of the ferroelectric ordering in novel smectic
 materials which have recently been investigated by Lemieux et.al.
 \cite{Lemieux1999, Lemieux2001, Lemieux2002, Lemieux2005}.
These materials, which are used as chiral dopants in a nonchiral
smectic C host, possess a molecular structure which is rather
different from that of conventional ferroelectric smectics.
 In particular, the molecular chirality is mainly determined by the
 chiral distribution of permanent dipoles within the molecular
 structure. It is shown that despite a different nature of
 intermolecular interactions, the spontaneous polarization in
 smectics C doped with such molecules is still determined by the
 same general mechanism.

  \section{Spontaneous polarization and order parameters of the
  smectic C$^*$ phase}

\subsection{Coupling between polarization, tilt and
chirality}\label{subsect21}

\begin{figure}
\centering\includegraphics[width=5cm]{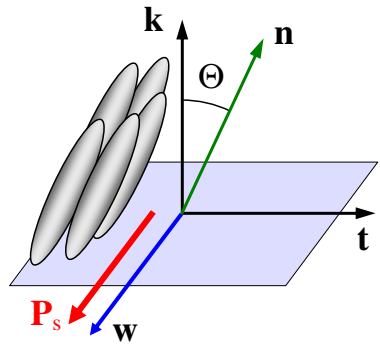} \caption{(Color
online) Spontaneous polarization $P_s$ and the pseudovector tilt
order parameter $w$ in the Sm{\it C}* phase.}\label{Fig3}
\end{figure}

It is well known since the work by R.~Meyer \cite{meyer} that from
the purely macroscopic point of view the ferroelectric ordering in
tilted smectics is determined by the linear coupling between the
polarization and the tilt in a chiral medium. In  terms of the
Landau-de Gennes expansion the free energy of the ferroelectric
smectic C$^*$ phase can be expressed as (see, for example,
\cite{ospikrev}):
\begin{equation}
F_C = F_A + F(\Theta) + \frac{1}{\chi_{\bot}} P_s^2 + c_p ({\bf P}_s
\cdot {\bf w}),
\end{equation}
where $F_A$ is the free energy of the smectic A phase, $F(\Theta)$
is the expansion of the excess free energy of the smectic C phase in
powers of the tilt angle $\Theta$ and the last two terms describe
the contribution which depends on the spontaneous polarization ${\bf
P}_s$. Here ${\bf w}= ({\bf n} \cdot {\bf k})({\bf k} \times {\bf
n})$ is the so-called pseudovector tilt order parameter of the
Sm{\it C} phase where ${\bf n}$ is the director and $ {\bf k}$ is
the smectic layer normal as shown in Fig.~\ref{Fig3}. Minimization
of the free energy (1) yields the well known result:
\begin{equation}
{\bf P}_s = c_p {\bf w} =c_p ({\bf n} \cdot {\bf k})({\bf k} \times
{\bf n}),
\end{equation}
which indicates that the spontaneous polarization is proportional to
the pseudovector tilt order parameter and the coupling constant
$c_p$.

One notes that the polarization is a polar vector while the tilt
order parameter ${\bf w}$ is a pseudovector with different
transformation properties (i.e, in contrast to the polar vector
${\bf P}_s$ it {\it does not} change  sign under space inversion).
Thus the linear relationship (2) between polarization and the tilt
is only possible if the coupling constant $c_p$ is a pseudoscalar
(which also changes sign under space inversion). Then the product of
the pseudoscalar $c_p$ and the pseudovector ${\bf w}$ makes the
polar vector like polarization. Pseudoscalar quantities are nonzero
only in a chiral medium. They are proportional to molecular
chirality and change sign when all chiral molecules reverse their
handedness. Thus one arrives at a well established conclusion that
the spontaneous polarization in tilted smectic phases occurs only if
at least a fraction of molecules are chiral.

At the same time, the pseudovector order parameter ${\bf w}$ is
nonzero also in the nonchiral smectic C phase. Indeed, ${\bf w}$ is
invariant under all symmetry transformations of the smectic C phase
including the reflection with respect to the tilt plane, which is a
symmetry plane. This is related to the transformation properties of
a pseudovector different from those of a polar vector. In this case,
${\bf w}$ is invariant under a reflection with respect to the tilt
plane because both vectors ${\bf n}$ and ${\bf k}$ are in the tilt
plane and thus are not effected by the reflection. In contrast, the
spontaneous polarization ${\bf P}_s$ , of course, changes sign under
a reflection with respect to the tilt plane. One notes that this
does not violate the linear relationship (2) because the
pseudoscalar parameter $c_p$ also changes sign under the reflection.
In a nonchiral smectic C phase the coupling constant $c_p$ vanishes
identically and the spontaneous polarization does not appear.

\subsection{Microscopic interpretation}\label{subsect22}

\begin{figure}
\centering\includegraphics[width=8cm]{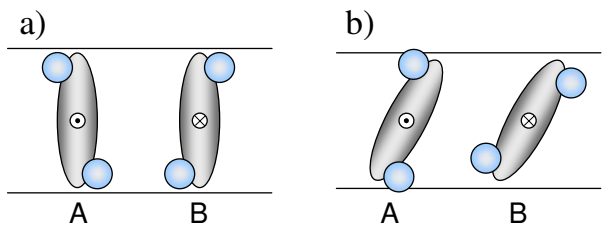} \caption{(Color
online) a) Molecular orientation A and B, which correspond to the
opposite directions of the molecular transverse dipole, are
equivalent in the Sm{\it A} phase. b) In the Sm{\it C}* phase
molecular orientation A is more energetically favorable than the
orientation B.}\label{Fig4}
\end{figure}

The purpose of any molecular theory of ferroelectric ordering in
tilted smectics is to establish a relationship between the general
macroscopic description presented in the previous subsection and the
molecular ordering on the microscopic level. An intuitive
interpretation of the ferroelectric ordering in the chiral smectic
C$^*$ phase \cite{boulder1,photinos1,giesselmann} can be illustrated
using Fig.~\ref{Fig4}.

For illustration only, let us consider a simple model of a chiral
biaxial molecule represented as a rigid rod with two 'lateral
groups' and a permanent dipole perpendicular to the molecular plane
as shown in Fig.~\ref{Fig4}a. Note that the lateral groups make the
molecule biaxial while the chirality is determined by the transverse
dipole. Without this dipole, the molecule possesses a mirror plane
and thus it is nonchiral. Now let us assume that the lateral groups
have a tendency to point in the direction of the region between two
adjacent smectic layers. One can readily see that in the smectic A
phase (i.e. without any tilt) the two orientations of such a
molecule, which correspond to opposite directions of the transverse
dipole ${\bm \mu}$ are energetically equivalent. Thus the
macroscopic polarization in the untilted smectic phase should
vanish. In contrast, in the tilted phase the balance between two
opposite directions of the transverse molecular dipole is violated
because the the molecular orientation A is more favorable than the
orientation B (see Fig.2b). As a result, the average molecular
dipole does not vanish, and a macroscopic polarization appears in
the direction perpendicular to the tilt plane. Now one has to
clarify how this type of ordering corresponds to the general
macroscopic description presented above.

Firstly one notes that this simple argument is valid for molecules
which are characterized by  the $C_{2h}$ symmetry when the
transverse dipole is removed. Simple examples of  molecules of the
$C_{2h}$ symmetry, presented in Fig.\ref{Fig1}, include a molecule
with two in-plane antiparallel dipoles (a model considered by
Goosens \cite{goosens1}) and a molecule of the 'zig-zag' shape
(considered by Wulf \cite{wulf} and others) which very roughly
characterize the actual shape of typical mesogenic molecules.
Orientation of any rigid biaxial molecule can be specified by the
unit vectors ${\bf a}$ and ${\bf b}$ in the direction of short and
long molecular axis, respectively. In practice, the orientation of a
nonpolar molecule  is usually characterized by second rank tensors
composed of the components of the vectors ${\bf a}$ and ${\bf b}$.
For example, uniaxial molecules are characterized by the molecular
tensor $Q^M_{\alpha \beta} = a_{\alpha}a_{\beta} - (1/3)
\delta_{\alpha \beta}$. The statistical average of ${\bf Q}^M$ is
the nematic tensor order parameter ${\bf Q}$. Biaxial molecules of
high symmetry, which possess two mutually perpendicular symmetry
planes, are also characterized by the second molecular tensor
$B^M_{\alpha \beta} =b_{\alpha}b_{\beta}- c_{\alpha}c_{\beta}$ where
the unit vector ${\bf c} \bot {\bf b}$ is the second molecular short
axis. In addition, the molecules of the $C_{2h}$ or lower symmetry
are characterized by the third molecular tensor
$a_{\alpha}b_{\beta}$ which is invariant under all symmetry
operations which leave the molecule intact. Indeed, one can readily
see that the molecules presented in Figs.\ref{Fig1} and \ref{Fig4}
are not invariant under the sign inversion of the axes ${\bf a}$ or
${\bf b}$ individually. At the same time, the molecules are
invariant under simultaneous inversion of both axis ${\bf a}$ and
${\bf b}$. This symmetry enables one to introduce the transverse
molecular pseudovector $({\bf a} \times {\bf b})$ which is related
to the skew part of $a_{\alpha}b_{\beta}$. One notes that the
existence of this transverse pseudovector does not violate the
mirror symmetry of the molecule because the pseudovector $({\bf a}
\times {\bf b})$ is invariant under a reflection with respect to the
molecular mirror plane which is parallel to ${\bf a}$ and ${\bf b}$.

Now it can readily be shown that the average $\langle{\bf a} \times
{\bf b}\rangle$ is nonzero only in a tilted smectic phase and is
proportional to ${\bf w} =({\bf n} \cdot {\bf k})({\bf k} \times
{\bf n})$ Thus the expression
\begin{equation}\label{wab}
{\bf w}_{ab} = \langle{\bf a} \times {\bf b}\rangle
\end{equation}
is a microscopic definition of a pseudovector tilt order parameter
of the smectic C phase. One notes that in the smectic C phase
composed of biaxial molecules there exist several tilt order
parameters. However, only the parameter (\ref{wab}) is directly
related to the spontaneous polarization.

Indeed, one notes that the molecular orientations A and B in
Fig.~\ref{Fig4} are characterized by the opposite directions of
molecular pseudovector $({\bf a} \times {\bf b})$ (because the the
short axis ${\bf b}$ points in the opposite directions while the
long axis ${\bf a}$ is the same). Thus one concludes that the
average pseudovector $\langle{\bf a} \times {\bf b}\rangle\neq 0$ in
the smectic C phase where the orientation A is more energetically
favorable than the orientation B. Moreover, it follows from the
general symmetry arguments that $\langle{\bf a} \times {\bf
b}\rangle \propto {\bf w} =({\bf n} \cdot {\bf k})({\bf k} \times
{\bf n})$ because ${\bf w}$ is the only pseudovector allowed by the
symmetry of the smectic C$^*$ phase. Indeed, any macroscopic vector
or pseudovector must be parallel to the $C_2$ symmetry axis of the
smectic C$^*$ layer which is normal to the tilt plane. This is
exactly the direction of ${\bf w}$. A different derivation of this
result is presented in Section 3.

Finally it can be shown that the spontaneous polarization ${\bf
P}_s$ is proportional to the average $\langle{\bf a} \times {\bf
b}\rangle \propto {\bf w}$. We assume for simplicity that the
transverse molecular dipole is parallel to the short molecular axis
${\bf c} \bot {\bf b} \bot {\bf a}$. i.e ${\bm \mu} = \mu_{\bot}
{\bf c}$. Then the macroscopic polarization in the smectic C$^*$
phase equals
\begin{equation}\label{Psmu}
{\bf P}_s = \rho \langle{\bm \mu}\rangle= \rho \mu_{\bot}
\langle{\bf c}\rangle,
\end{equation}
where $\rho$ is the molecular number density.

The unit vector ${\bf c}$ can be expressed in terms of the unit
vectors ${\bf a}$ and ${\bf b}$ in the following way
\begin{equation}
{\bf c} = \Delta ({\bf a} \times {\bf b}),
\end{equation}
where $\Delta =(({\bf a} \times {\bf b}) \cdot {\bf c})$ is the
molecular unit pseudoscalar which specifies the handedness of the
molecular coordinate system. Note that in Eq.(5) ${\bf c}$ is the
conventional polar vector which is expressed as a product of the
pseudovector $({\bf a} \times {\bf b})$ and the pseudoscalar
$\Delta$.

Now the spontaneous polarization (4) can be rewritten as:
\begin{equation}\label{PsDelta}
{\bf P}_s =\rho \mu_{\bot} \langle{\bf c}\rangle=\\
\rho \mu_{\bot} \Delta \langle{\bf a} \times {\bf b}\rangle= \rho
\Delta_{\mu} \langle{\bf a} \times {\bf b}\rangle,
\end{equation}
 where $\Delta_{\mu} = ({\bm \mu} \cdot ({\bf a} \times {\bf
b}))$. Here we have taken into account that the parameter $\Delta$
is independent of the molecular orientation.

According to Eq.(6), the spontaneous polarization ${\bf P}_s$ is
proportional to the  tilt order parameter $\langle{\bf a} \times
{\bf b}\rangle \propto {\bf w} =({\bf n} \cdot {\bf k})({\bf k}
\times {\bf n})$. Thus the simple microscopic interpretation of the
appearance of the spontaneous polarization in the smectic C$^*$
phase, presented in this subsection, is fully consistent with the
general phenomenological theory outlined in
subsection~\ref{subsect21}.

In summary, one concludes that if the smectic C$^*$ phase is
composed of molecules of the $C_{2h}$ symmetry or lower with an
additional transverse dipole in the direction perpendicular to the
molecular plane, the spontaneous polarization is proportional to the
order parameter (\ref{wab}) and the pseudoscalar quantity
$\Delta_{\mu}=({\bm \mu} \cdot ({\bf a} \times {\bf b}))$ which
characterizes the molecular chirality determined by the transverse
dipole ${\bm \mu}$. A consistent statistical theory of the
ferroelectric ordering in the smectic C$^*$ phase is presented in
the following section.

\section{Molecular statistical theory of ferroelectric ordering in
the smectic C$^*$ phase}

\subsection{General results}

As discussed in the previous section, the orientation of a biaxial
molecule can be specified by the unit vectors ${\bf a}$ and ${\bf
b}$ in the direction of the long and short molecular axis,
respectively. The second molecular short axis ${\bf c}$ is then
given by Eq.(5). The spontaneous polarization can be expressed in
the following general form:
\begin{equation}\label{Ps}
{\bf P}_s = \rho \langle{\bm \mu}_{\bot}\rangle = \rho \int {\bm
\mu}_{\bot} f_1 ({\bf a}, {\bf b}) d {\bf a} d {\bf b},
\end{equation}
where $f_1 ({\bf a}, {\bf b})$ is the orientational distribution
function of the smectic C$^*$ phase, and the transverse molecular
dipole ${\bm \mu}_{\bot} =  \mu_{\bot} {\bf c}$. The orientational
distribution function can always be expressed in terms of the
effective one-particle potential $U_1 ({\bf a}, {\bf b})$:
\begin{equation}\label{f1}
f_1 ({\bf a}, {\bf b}) = \frac{1}{Z} \exp \left[ -  \frac{U_1 ({\bf
a}, {\bf b})}{k_BT} \right],
\end{equation}
where
\begin{equation}\label{Z}
Z = \int \exp \left[ -  \frac{U_1 ({\bf a}, {\bf b})}{k_BT} \right]
d {\bf a} \ d {\bf b}.
\end{equation}

For molecules which are polar in the direction of the ${\bf
c}$-axis, in the quadrupolar approximation the effective
one-particle potential depends on the unit vector ${\bf c}$ and
the second rank molecular tensors $a_{\alpha}a_{\beta},
b_{\alpha}b_{\beta}$ and $a_{\alpha}b_{\beta}$ which are invariant
under all symmetry operations of a molecule. Here ${\bf c}$ is
expressed in terms of ${\bf a}$ and ${\bf b}$ by Eq.(5). As
discussed in subsection \ref{subsect22}, the invariant
$a_{\alpha}b_{\beta}$ exists only for biaxial molecules of
sufficiently low symmetry. Thus in the quadrupolar approximation
the potential $U_1 ({\bf a}, {\bf b})$ reads:
\begin{multline}\label{U1}
U_1 ({\bf a}, {\bf b})= a_{\alpha}a_{\beta}A_{\alpha \beta} +
b_{\alpha}b_{\beta} B_{\alpha \beta} +\\ (a_{\alpha}b_{\beta}+
b_{\alpha}a_{\beta})C_{\alpha \beta} + ({\bf a}\times\bf{b})\cdot
{\bf W},
\end{multline}
where  ${\bf W}$ and $A_{\alpha \beta}, B_{\alpha \beta}, C_{\alpha
\beta}$ are the material pseudovector and tensors correspondingly,
which depend on the symmetry of the chiral smectic C$^*$ phase. This
symmetry, in turn, is determined by the two tensors
$n_{\alpha}n_{\beta}$ and $k_{\alpha}k_{\beta}$ which specify the
macroscopic structure of the phase. One notes that the last term in
Eq.(10) contains a pseudovector $({\bf a} \times {\bf b})$, which
means that ${\bf W}$ has also to be a pseudovector in order to
insure that $U_1({\bf a}, {\bf b})$ is a scalar. We conclude that
this last term is determined by chiral interactions in the system,
i.e., the interactions depending on the handedness of the
interacting molecules. Indeed, let us consider the case when Eq.(10)
describes an effective one-particle potential of a chiral dopant in
the chiral smectic C* host. Inversion of the host chirality results
in sign inversion of the pseudovector ${\bf W}$ while all material
tensors $A_{\alpha \beta}, B_{\alpha \beta}$ and $C_{\alpha \beta}$
remain the same. As a result, the last term in Eq.(10) changes sign.
Thus the last term describes the so called chiral discrimination
energy, i.e. the difference of energy of interaction between the
same chiral molecule and the two enantiomeric forms of the chiral
host. In many cases such a discrimination is small and then this
term  may be neglected.

Taking into account that all macroscopic quantities in the smectic C
phase must be quadratic both in ${\bf n}$ and ${\bf k}$, the
pseudovector ${\bf W}$ can be expressed in the following general
form:
\begin{equation}
{\bf W} = \omega {\bf P}_s + \kappa {\bf w},
\end{equation}
where $\omega$ is a pseudoscalar and $\kappa$ is a scalar. The
symmetric tensors ${\bf A}$, ${\bf B}$ and ${\bf C}$ are expressed
as:

\[
A_{\alpha \beta} = A_1 n_{\alpha}n_{\beta} + A_2 k_{\alpha}k_{\beta}
+ A_3({\bf n\cdot k})( n_{\alpha}k_{\beta}+ k_{\alpha}n_{\beta}),
\]
\[
B_{\alpha \beta} = B_1 n_{\alpha}n_{\beta} + B_2 k_{\alpha}k_{\beta}
+ B_3({\bf n\cdot k})( n_{\alpha}k_{\beta}+ k_{\alpha}n_{\beta}),
\]
\begin{equation}
C_{\alpha \beta} = C_1 n_{\alpha}n_{\beta} + C_2 k_{\alpha}k_{\beta}
+ C_3({\bf n\cdot k})( n_{\alpha}k_{\beta}+ k_{\alpha}n_{\beta}),
\end{equation}
where we have neglected the terms quadratic in ${\bf P}_s$ and ${\bf
w}$ because these terms are of the order of $\Theta^2$ at small tilt
angle $\Theta \ll 1$ while all other terms in Eq.(12) are of the
order of $1$.

The effective one particle potential $U_1 ({\bf a}, {\bf b})$ is
determined  by intermolecular interactions in the smectic C$^*$
phase which may, or may not be sensitive to molecular chirality. For
example, the electrostatic interaction between permanent dipoles and
quadrupoles is not sensitive to molecular chirality while the
interactions involving molecular octupoles are different for the
pairs of molecules of equal and opposite handedness, respectively.
In particular, the pseudovector quantity $\bf W$ in Eq.(11) must
vanish if the molecules are nonchiral. This parameter is determined
by some chiral intermolecular interactions which exist only between
chiral molecules. All other material parameters in Eqs.(10-12) are
scalars, and therefore they are generally nonzero in the
corresponding nonchiral smectic C phase. The difference between
scalar and pseudoscalar material parameters in the effective
one-particle potential enables one to distinguish between two
different microscopic mechanisms of ferroelectric ordering in the
smectic C$^*$ phase.

Taking into account that the spontaneous polarization ${\bf P}_s$
and the pseudovector ${\bf w}$ are small at small tilt angles
$\Theta$, the orientational distribution function (8) can be
expanded in powers of ${\bf P}_s$ and ${\bf w}$ keeping the linear
terms, and substituted into the general expression for the
spontaneous polarization (7). As a result, the spontaneous
polarization in the smectic C$^*$ phase can be expressed as:
\begin{multline}
\bm \xi \cdot {\bf P}_s = \rho \Delta \kappa {\bm \lambda} \cdot
{\bf w} +\\ \rho \int {\bm \mu}_{\bot} \frac{1}{Z_0} \exp \left[ -
\frac{U_1^{(0)} ({\bf a}, {\bf b})}{k_BT} \right]\ d{\bf a}\ d{\bf
b},
\end{multline}
where the inverse polarizability tensor $\xi_{\alpha \beta} =
\delta_{\alpha \beta}+ (\rho/k_BT)\langle\omega
\mu_{\bot\alpha}c_{\beta}\rangle_0$, the tensor $\lambda_{\alpha
\beta} =-(\rho/k_BT)\langle \mu_{\bot\alpha}c_{\beta}\rangle_0$ ,
the averaging $\langle...\rangle_0 $ is performed with the
orientational distribution function $f_0 = (1/Z_0)
\exp(-U_1^{(0)}({\bf a}, {\bf b})/k_BT)$ and $U_1^{(0)}({\bf a},
{\bf b})$ is given by Eqs.(10-12) with ${\bf W}=0$.

According to Eq.(13) there exist two qualitatively different
contributions to the spontaneous polarization of the smectic C$^*$
phase which correspond to the two terms in the r.h.s. of Eq.(13).
The first term comes from the last term in the one-particle
potential (\ref{U1}), which is determined by chiral intermolecular
interactions. This contribution corresponds to the microscopic
mechanism of ferroelectric ordering determined by chiral
interactions which has been considered in
refs.\cite{osipov84,ospikrev,MCLCreview,ossteg}. In contrast, the
second term describes the ferroelectric ordering of chiral molecules
determined by nonchiral intermolecular interactions. Indeed, this
contribution describes the ordering of a transverse molecular dipole
in a nonchiral effective one-particle potential $U_1^{(0)}({\bf a},
{\bf b})$. This potential does not depend on any pseudoscalar
parameters, and thus, in the first approximation, it is exactly the
same as in the corresponding nonchiral smectic C phase. Molecular
chirality in this case is determined by the orientation of the
molecular transverse dipole and manifests itself during the
averaging process as described below. Now let us consider this
contribution in more detail.

In the general case the effective one-particle potential
$U_1^{(0)}({\bf a}, {\bf b})$ can be expressed as a sum of the
following two terms:
\begin{equation}
U_1^{(0)} ({\bf a} , {\bf b} ) = U_a ({\bf a}) + U_b ({\bf a} , {\bf
b} ) ,
\end{equation}
where $U_a({\bf a}) =a_{\alpha}a_{\beta}A_{\alpha \beta} $ and $U_b
({\bf a}, {\bf b})=b_{\alpha}b_{\beta} B_{\alpha \beta} +
(a_{\alpha}b_{\beta}+ b_{\alpha}a_{\beta})C_{\alpha \beta}$. In
Eq.(14) the first term depends only on the orientation of the long
molecular axis ${\bf a}$ while the second term depends also on the
orientation of the short axis ${\bf b}$. Smectic liquid crystals are
composed of strongly anisotropic molecules, and in this case one may
assume that the intermolecular interaction energy associated with a
change of orientation of the long axes is much larger than the
change of the energy associated with  the rotation of short
molecular axes. Then the second term in Eq.(14) is expected to be
much smaller than the first one, i.e. $U_b \ll U_a$. One notes that
the first term in Eq.(14) determines the orientational (nematic)
ordering of long molecular axis, while the second term is
responsible for a weaker ordering of short molecular axes. Now the
orientational distribution function $f_0 ({\bf a} , {\bf b} )$ can
be expanded in powers of $U_b$ keeping the linear term:
\begin{equation}
\frac{1}{Z_0} \exp\left[-\frac{U_1^{(0)}(\bf a, b) }{k_BT}\right]
\approx f_a ({\bf a}) \left[1- \frac{U_b ({\bf a} , {\bf b})}{k_BT}
\right],
\end{equation}
where the uniaxial orientational distribution function $f_a ({\bf
a})$ depends only on the orientation of the long axis ${\bf a}$:
\begin{equation}
f_a ({\bf a} ) = \frac{1}{Z_a } \exp\left[-\frac{U_a ({\bf a})
}{k_BT}\right].
\end{equation}
Substituting Eq.(15) into the second term in  Eq.(13) one obtains
the following expression for the spontaneous polarization
\begin{equation}
{\bf P}_s \approx - \Delta \frac{\rho \mu_{\bot}}{k_BT} \int ({\bf
a\times b})\ U_b ({\bf a} , {\bf b} )\  f_a ({\bf a} ) d {\bf a} d
{\bf b}.
\end{equation}
Now Eqs.(14) and (16) can be substituted into Eq.(17) where the
averaging is performed over  ${\bf b}$ and ${\bf a}$ taking into
account that the function $f_a ({\bf a})$ is independent of ${\bf
b}$. Neglecting biaxiality of the smectic C phase one may use the
following simple formulae:
\begin{equation}\label{bav}
\frac{1}{2\pi} \int b_{\alpha}b_{\beta} d {\bf b} =
\frac{1}{2}(\delta_{\alpha \beta} -a_{\alpha}a_{\beta});
\end{equation}
 and
\begin{equation}\label{aav}
\frac{1}{4\pi} \int f_a ({\bf a})(a_{\alpha}a_{\beta} - \frac{1}{3}
\delta_{\alpha \beta})\ d {\bf a} = S(n_{\alpha}n_{\beta} -
\frac{1}{3}\delta_{\alpha \beta}),
\end{equation}
and obtain the final expression for the spontaneous polarization:
\begin{equation}\label{Psgen}
{\bf P}_s = \frac{\rho \Delta_{\mu} C_3 S}{2 k_BT}\  [{\bf n} \times
{\bf k}]({\bf n} \cdot {\bf k}),
\end{equation}
where $S$ is the nematic order parameter.

 Eq.(19), which has been obtained without using any particular model, presents a
general expression for the spontaneous polarization of the chiral
smectic C$^*$ phase determined by nonchiral intermolecular
interactions. The spontaneous polarization is proportional to the
pseudoscalar parameter $\Delta_{\mu} = ({\bm \mu}_{\bot} ({\bf a}
\times {\bf b}))$ which specifies molecular chirality determined by
the orientation of the transverse dipole, and the parameter $C_3$
which is determined by nonchiral interactions. One notes that the
general microscopic mechanism of ferroelectric ordering described by
the present theory is qualitatively the same as the one considered
in the Boulder model \cite{boulder1,boulder2}. In the framework of
the Boulder model the restrictions imposed by the binding cite on
the orientation of a given molecule are equivalent to the specific
form of the effective one-particle potential $U_1^{(0)}({\bf a},
{\bf b})$. From the qualitative point of view, both the binding cite
and the effective one-particle potential possess the same symmetry
as the nonchiral smectic C phase (i.e. the $C_2$ symmetry axis and
the mirror plane) and are not sensitive to molecular chirality. On
the other hand, the present theory also covers the model of Goosens
\cite{goosens2} based on the electrostatic interaction between
molecules with pairs of antiparallel dipoles. Detailed numerical
results obtained using a version of this model are presented in
Section~\ref{section4}. Explicit expressions for effective
one-particle potential and the spontaneous polarization in the
quadrupole-quadrupole interaction model are derived in the following
section using the mean-field approximation.

\subsection{Quadrupole-Quadrupole interaction model}

In this section we obtain explicit analytical results for the
spontaneous polarization in the smectic C$^*$ phase using the simple
model of a rigid molecule with essentially uniaxial quadrupole
tensor slightly tilted with respect to the primary molecular axis
(which is determined by the shape or molecular inertia tensor and
orders along the director). Such a molecule is biaxial, and the
molecular biaxiality is determined by the angle between the axis of
the quadrupole tensor and the primary molecular axis. Assuming that
this angle is small, one obtains approximate expressions for the
polarization using the generalized mean-field theory.

Let us consider the system of rigid molecules with primary axes
${\bf a}_i$. The molecules are  characterized by the permanent
quadrupole tensor $q_{\alpha \beta}$ which depends on the
distribution of effective charges within a molecule. It should be
noted that for molecules of the $C_{2h}$ symmetry, considered here,
one primary axis of any molecular tensor (including the quadrupole
one) must be parallel to the two-fold symmetry axis of the molecule
${\bf c}$. At the same time, the orientation of the two remaining
primary axes, which lie in the symmetry plane of the molecule, is
not specified by the molecular symmetry. As a result, the
orientation of these axes should generally be different for
different molecular tensors characterizing different molecular
properties. In particular, the primary axes of the quadrupole tensor
$q_{\alpha \beta}$ are not expected to coincide with those of the
molecular inertia tensor, ${\bf a}$ and ${\bf b}$. Thus in the
general case the traceless symmetric quadrupole tensor can be
expressed in terms of molecular axes ${\bf a} \bot {\bf b} \bot {\bf
c}$ in the following way:
\begin{equation}\label{q}
q_{\alpha \beta} = q_1( a_{\alpha} a_{\beta} -
\delta_{\alpha\beta}/3)+ q_2 (b_{\alpha}b_{\beta} -c_{\alpha}
c_{\beta}) + q_3 (a_{\alpha} b_{\beta} +b_{\alpha} a_{\beta}).
\end{equation}
It will be shown below that the spontaneous polarization is
proportional to the off-diagonal element $q_3$, which characterizes
the difference in the orientation of the molecular inertia and
quadrupole tensors.

The electrostatic interaction between permanent quadrupoles of the
molecules $i$ and $j$ can be written in the form:
\begin{equation}\label{Uqq}
U_{qq}(i,j) =  q^i_{\alpha \beta} D_{\alpha \beta \gamma \delta}
q^j_{\gamma \delta},
\end{equation}
where the quadrupole-quadrupole coupling tensor ${\bf D}$ is given
by
\begin{multline}\label{Dqq}
D_{\alpha \beta \gamma \delta}=  \frac{3}{4R^5 }( \delta_{\alpha
\delta}\delta_{\beta \gamma} +\delta_{\alpha \beta}\delta_{\gamma
\delta} + \delta_{\alpha \gamma}\delta_{\beta \delta}
-\\5\delta_{\alpha \beta} u_{\gamma}u_{\delta} -5\delta_{\alpha
\gamma} u_{\beta}u_{\delta} -5\delta_{\alpha \delta}
u_{\gamma}u_{\beta} -5\delta_{\beta \gamma} u_{\delta}u_{\alpha}
-\\5\delta_{\delta \beta} u_{\gamma}u_{\alpha} -5\delta_{\gamma
\delta} u_{\alpha}u_{\beta} +35 u_{\alpha}u_{\beta}
u_{\gamma}u_{\delta}),
\end{multline}
where ${\bf R}={\bf R}_{ij}$ is the intermolecular vector and the
unit vector ${\bf u}= {\bf R}/|{\bf R}|$.

Now the quadrupole-quadrupole interaction can be taken into account
in the generalized mean-filed approximation of the smectic C$^*$
phase. In this approximation (see, for example
\cite{straley,gelbart,book}) the free energy of the system without
positional order can be expressed as:
\begin{multline}\label{Fmf}
F/V= \frac{1}{2} \rho^2 \int f_1(\Omega_1) U(1,2) f_1 (\Omega_2)
d^2{\bf R}\  d \Omega_1 d \Omega_2 +\\\rho k_BT \int f_1 (\Omega_1)
\ln f_1 (\Omega_1) \ d \Omega_1,
\end{multline}
where $f_1 (\Omega)$ is the one-particle orientational distribution
function, the variable $\Omega_i = [{\bf a}_i, {\bf b}_i]$ specifies
the orientation of the molecule $i$, $U(1,2)$ is the effective pair
interaction potential which takes into account the steric cut-off,
and $\rho$ is the number density of molecules per unit area of the
smectic layer.

Minimization of the free energy (24) yields the one-particle
distribution function in the form (\ref{f1}, \ref{Z}) with the
one-particle potential $U_1$ being equal to the mean-field potential
\begin{equation}\label{UMFdef}
U_{MF}({\bf a}, {\bf b})= \rho \int U(1,2) f_1({\bf a}_2, {\bf b}_2)
d^2 {\bf R} \  d {\bf b}_2\  d {\bf a}_2.
\end{equation}

The total effective pair interaction potential $U(1,2)$ can be
expressed as
\begin{equation}\label{U12_0_qq}
U(1,2) = U_{0}(1,2) + U_{qq}(1,2),
\end{equation}
where $U_{0}(1,2)$ is the effective interaction potential for
molecules without permanent quadrupole moments, and $U_{qq}(1,2)$ is
the quadrupole-quadrupole interaction energy given by
Eq.(\ref{Uqq}). We assume for simplicity that the potential
$U_0(1,2)$ is even in ${\bf a}_1, {\bf a}_2, {\bf b}_1, {\bf b}_2,
$, and thus it cannot be responsible for the ferroelectric ordering
in the smectic C$^*$ phase.

Substituting Eqs.(\ref{Uqq},\ref{Dqq}) for the quadrupole-quadrupole
interaction potential into Eq.(\ref{U12_0_qq}) and then into
Eq.(\ref{UMFdef}) one obtains the following expression for the total
mean-field potential after averaging over ${\bf a}_2, {\bf b}_2$ and
integration over ${\bf R}$:
\begin{equation}
U_{MF}({\bf a}, {\bf b})= U^{(0)}_{MF}({\bf a}, {\bf b}) +
U^{qq}_{MF}({\bf a}, {\bf b}),
\end{equation}
where $ U^{(0)}_{MF}({\bf a}, {\bf b})$ is even in ${\bf a}$ and
${\bf b}$, and where
\begin{equation}
U^{qq}_{MF}({\bf a}, {\bf b})=\frac{\pi}{4} \rho R^2_0\
q^{(1)}_{\alpha \beta}\  D_{\alpha \beta \gamma \delta}({\bf k})
\langle q_{\gamma \delta}\rangle.
\end{equation}
Here $q^{(1)}_{\alpha \beta}$ is the quadrupole tensor of the
molecule $1$ given by Eq.(\ref{q}), $R_0$ is the distance of minimum
approach between the neighboring molecules in the layer, and the
tensor ${\bf D}$ is given  by Eq.(\ref{Dqq}) with $R=R_0$ and ${\bf
u}={\bf k}$.

At this point we can neglect the weak biaxiality of the molecular
distribution and use the average of the quadrupole tensor (\ref{q})
in the simple form
\begin{equation}
\langle q_{\alpha \beta}\rangle = q_1 S\ (n_{\alpha}n_{\beta} -
\delta_{\alpha \beta}/3),
\end{equation}
After some algebra the corresponding expression for $ D_{\alpha
\gamma \delta \beta}({\bf k}) \langle q_{\gamma \delta}\rangle$ can
be written as:
\begin{multline}
M_{\alpha \beta} =D_{\alpha \beta \gamma \delta}({\bf k}) \langle
q_{\gamma \delta}\rangle= \frac{3}{4R^5_0}\  q_1 S \times\\ [
\delta_{\alpha \beta}(1 -5 \cos^2 \Theta) + 2 n_{\alpha}n_{\beta}
-5(1 - 7 \cos^2 \Theta )k_{\alpha}k_{\beta}-\\10({\bf k} \cdot {\bf
n}) (n_{\alpha}k_{\beta}+n_{\beta}k_{\alpha}) ].
\end{multline}

Assuming that the off-diagonal element $q_{3}$ of the molecular
quadrupole tensor is small one may expand the exponent in
Eqs.(\ref{f1},\ref{Z}) in powers of $q_3$ keeping the linear
terms. This yields the following approximate expression for the
orientational distribution function:
\begin{equation}\label{f1expand}
f_1 ({\bf a}, {\bf b}) =f_0 ({\bf a}, {\bf b}) + \Delta f ({\bf a},
{\bf b}),
\end{equation}
where
\begin{equation}\label{f0}
f_0 ({\bf a}, {\bf b}) = \frac{1}{Z_0} \exp\left[ -\frac{
U^{(0)}_{MF}({\bf a}, {\bf b})}{k_BT}+\frac{ \pi \rho R_0^2
q^0_{\alpha \beta} M_{\alpha \beta}}{4 k_BT}\right],
\end{equation}
and
\begin{equation}\label{Deltaf}
\Delta f ({\bf a}, {\bf b}) = \frac{\pi \rho}{4 k_BT} R_0^2
q_3(a_{\alpha}b_{\beta} + a_{\beta}b_{\alpha})M_{\alpha \beta}\
f_0({\bf a}, {\bf b}),
\end{equation}
and where $q_{\alpha \beta}^0$ is the molecular quadrupole tensor
given by Eq.(\ref{q}) without the off-diagonal term.

Only the term containing $\Delta f$ makes a contribution to the
spontaneous polarization given by the general equations
(\ref{PsDelta},\ref{Ps}). Substituting
Eqs.(\ref{f1expand}-\ref{Deltaf}) and assuming that the biaxial
ordering in the system is weak (i.e. using
Eqs.~(\ref{bav},\ref{aav}) for the averaging) one obtains the
following expression for the spontaneous polarization of the smectic
C$^*$ phase:
\begin{equation}\label{Pq}
{\bf P}_s = \rho \Delta_{\mu} c^0_p [{\bf n} \times {\bf k}]({\bf n}
\cdot {\bf k}),
\end{equation}
with
\begin{equation}
c^0_p = \frac{15\pi \rho}{16 k_B T R^3_0}\  q_1q_3 S^2 (
4-7\sin^2\Theta).
\end{equation}

This explicit analytical expression for the spontaneous polarization
was obtained in the context of the model of the
quadrupole-quadrupole intermolecular interactions and confirms all
results of the  previous section which have been obtained using
general theory arguments. In particular, the spontaneous
polarization is proportional to the pseudoscalar parameter
$\Delta_{\mu}=({\bm \mu} \cdot [{\bf a} \times {\bf b}])$ which
specifies molecular chirality. At the same time, the remaining
factor $c_p^0$ is completely independent of the molecular chirality,
and is obtained using the orientational distribution function of the
nonchiral smectic C phase. In this simple model, the factor $c_p^0$
depends only on the diagonal components of the molecular quadrupole
moment, the orientational order parameters of the smectic C phase
and the distance of minimum approach $R_0$.

In the following section we consider a more realistic model based
on electrostatic interactions between localized molecular dipoles.

\section{Interaction between chiral pairs of molecular
dipoles}\label{section4}

\begin{figure}
\centering\includegraphics[width=4cm]{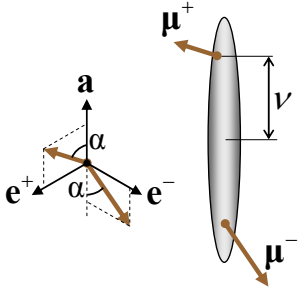} \caption{Schematic
of molecular axes and dipoles}\label{Fig5}
\end{figure}

In  a recent series of papers  by Lemieux et.al. \cite{Lemieux1999,
Lemieux2001, Lemieux2002, Lemieux2005} a number of novel compounds
with unconventional structure have been used as chiral dopants to
induce the large spontaneous polarization in the smectic C phase.
One notes that the chirality of these molecules is not determined by
any chiral centers, but is a consequence of a chiral distribution of
permanent molecular dipoles. Disregarding other elements of the
actual molecular structure, one can use the minimum model shown in
Fig.~\ref{Fig5}. In this model, the molecule is presented by a rigid
rod (with some dispersion interactions between rods) and a pair of
off-center dipoles with large transverse components lying in
orthogonal planes. Introducing the orthogonal transverse unit
vectors $\bf e^\pm$ we write
\begin{equation}\label{mupm}
\bm \mu^\pm =\mu({\bf e}^\pm\sin\alpha\pm{\bf a}\cos\alpha).
\end{equation}
One can readily see that the molecule presented in Fig.~\ref{Fig5}
is chiral because it does not have any symmetry planes. The total
dipole moment of the molecule is transverse $\bm\mu_\bot=(\bm
\mu^++\bm \mu^-)$ and directed along the unit vector ${\bf c=(\bf
e^++e^-)}/\sqrt{2}$. Accordingly, another short molecular axis is
to be defined as ${\bf b=(\bf e^+-e^-)}/\sqrt{2}$. The spontaneous
polarization in such a system is given by the general eqs.
(\ref{Psmu},\ref{PsDelta}) where the pseudoscalar parameter is
expressed as:
\begin{equation}
\Delta_\mu=\frac{\sqrt{2}}{\mu\sin\alpha}\ {\bf a}\cdot[\bm
\mu^+\times\bm \mu^-]=\sqrt{2}\Delta\ \mu\sin\alpha,
\end{equation}
This parameter quantitatively determines the molecular chirality
in the context of the present model..

An interaction potential for a pair of such molecules is expressed
as a sum of effective interaction potentials between rigid uniaxial
cores and the sum of all electrostatic dipole-dipole interactions:
\begin{equation}
U(1,2) = U^{\rm aa} ({\bf a}_1, {\bf R}, {\bf a}_2) + U_{\mu}(1,2)
\end{equation}
with $U_\mu$ depending also on the orientation of short molecular
axes:
\begin{multline}\label{U12}
U_\mu(1, 2)={\bm \mu}_1^+ \cdot {\bf D}_{\uparrow\uparrow}\cdot{\bm
\mu}_2^+ +{\bm \mu}_1^- \cdot {\bf D}_{\uparrow\uparrow}\cdot{\bm
\mu}_2^- \\+{\bm \mu}_1^+ \cdot {\bf
D}_{\uparrow\downarrow}\cdot{\bm \mu}_2^- +{\bm \mu}_1^- \cdot {\bf
D}_{\uparrow\downarrow}\cdot{\bm \mu}_2^+,
\end{multline}
where the both tensors $\bf D$ involved have the form
\begin{equation}\label{Ddd}
D_{ij}({\bf a}_1, {\bf a}_2, {\bf R})= \frac{1}{r^5}\left(r^2
\delta_{ij}-3 r_i r_j\right),
\end{equation}
with the distance between the interacting dipoles $\bf r$ being the
function of molecular orientation and intermolecular distance: ${\bf
r}={\bf r}({\bf a}_1, {\bf a}_2, {\bf R})$. For ${\bf
D}_{\uparrow\uparrow}$ it reads ${\bf r}={\bf R}+\nu {\bf
a}_2-\nu{\bf a}_1$, while for ${\bf D}_{\uparrow\downarrow}$ it is
${\bf r}={\bf R}+\nu {\bf a}_2+\nu{\bf a}_1$.

We assume that the tilting of the director in the smectic C phase is
due to the long-axes potential $U^{\rm aa}(1,2)$. We then employ the
following model expression for the uniaxial potential which has been
extensively used in  the general theory of Sm{\it A}--Sm{\it C}
transition \cite{uniaxial1, uniaxial2}:
\begin{multline} \label{Uaa}
U^{\rm aa} ({\bf a}_1, {\bf R}, {\bf a}_2) \approx v_1(R)
\left[({\bf a}_1 \cdot {\bf u})^2+({\bf a}_2 \cdot {\bf u})^2\right]
\\
+v_2(R) ({\bf a}_1\cdot {\bf a}_2)^2+v_3(R)({\bf a}_1\cdot{\bf
a}_2)({\bf a}_1\cdot{\bf u})({\bf a}_2\cdot{\bf u}) \\+ v_4(R)({\bf
a}_1\cdot{\bf u})^2({\bf a}_2\cdot{ \bf u})^2,
\end{multline}
As discussed in detail in \cite{uniaxial1, uniaxial2} the
corresponding mean-field potential
\begin{multline}\label{Umfaa}
U_{\rm MF}^{\rm aa}({\bf a})= w_1 P_2(\cos \gamma)+w_2 S_k P_2(\cos
\gamma)  +\\w_3 P_k \sin^2 \gamma \cos 2\phi + w_4 V \sin 2\gamma
\cos \phi,
\end{multline}
depends on three order parameters
\begin{eqnarray}\label{SkPk}
S_k =\langle P_2(\cos \gamma)\rangle,\ \  P_k=\langle\sin^2 \gamma \cos 2\varphi\rangle,\ \\
V=\langle\sin 2\gamma \cos \varphi\rangle\label{C},
\end{eqnarray}
where $\gamma$ and $\varphi$ are the polar and azimuthal angles of
the unit vector $\bf a$ respectively.

If these order parameters are known, conventional order parameters
such as nematic order parameter $S$, nematic tensor biaxiality $P$,
and the tilt angle $\Theta$ can be easily calculated as
\begin{eqnarray}
\tan 2\Theta =\frac{V}{S_k -0.5 P_k}, \label{Theta}\\
S=\frac{1}{4}S_k + \frac{3}{8}P_k+\frac{3\ V}{4 \sin 2\Theta},\label{S}\\
B=\frac{1}{2}S_k + \frac{3}{4}P_k-\frac{V}{2 \sin 2\Theta}\label{P}.
\end{eqnarray}
The potential (\ref{Umfaa})promotes the tilt if the nematic order
parameter exceeds the critical value
\begin{equation}\label{SAC}
S_{AC}=\frac{3w_1}{4w_4-3w_2},
\end{equation}
which means that the growth of the nematic order is the driving
force of the tilting transition. Thus in this model the molecular
dipoles are not responsible for the tilt of the director, but the
interaction between such pairs of dipoles gives rise to the
spontaneous polarization as shown below.

One notes that the actual form of the interaction potential for
off-center dipoles is too complicated to be used directly in the
statistical theory \cite{uniaxial2}. In particular, substituting the
actual potential (\ref{U12}) into the Eq.(\ref{UMFdef}) one cannot
obtain the mean-field potential as an explicit function of the
orientational order parameters. As a result the free energy cannot
be minimized to determine the transition point and the temperature
variation of the parameters. However, we can expand the actual
dipole-dipole potential in spherical invariants neglecting the
higher order terms. One notes that the statistical averages of
higher order terms are expressed in terms of higher order
orientational order parameters which are normally not important from
the qualitative point of view \cite{book}. The details of the
expansion procedure are presented in the Appendix~\ref{appendix}.

As shown in the Appendix~\ref{appendix}, the actual interaction
potential between the pairs of off-center dipoles can be
approximated by the relatively simple expression:
\begin{multline}\label{UMFTc}
U_{\rm MF}^{\rm \Gamma}=w_5[\cot\alpha\ V({\bf a}\cdot {\bf k})({\bf
b}\cdot {\bf t})+\\\cot\alpha\ \Gamma\ \sin2\gamma\cos\varphi\
+\sqrt{2}\ \Gamma\ ({\bf a}\cdot {\bf k})({\bf b}\cdot {\bf t})],
\end{multline}
which involves the biaxial order parameter $\Gamma=\langle({\bf
a}\cdot {\bf k})({\bf b}\cdot {\bf t})\rangle$. Combining this with
the uniaxial mean-field potential (\ref{Umfaa}) we can write the
total mean-filed approximation of the LC free energy (\ref{Fmf}).

\begin{figure}
\centering\includegraphics[width=7.5cm]{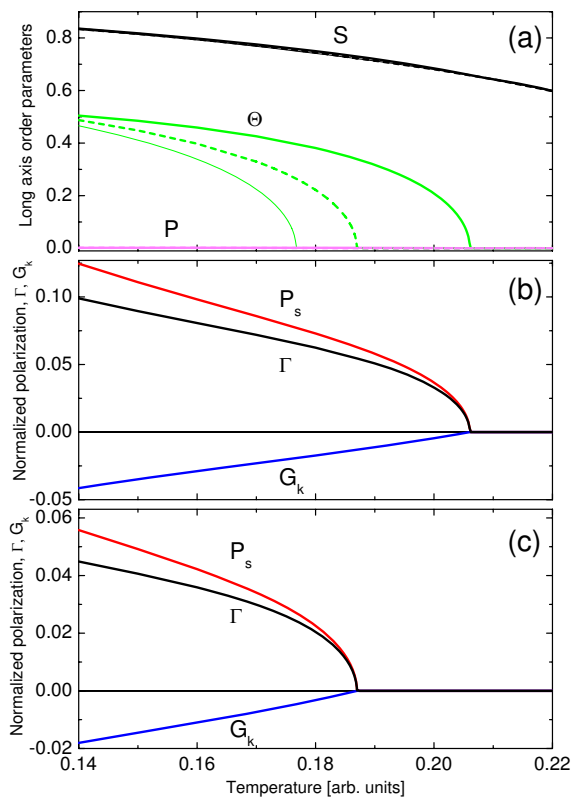} \caption{(Color
online) (a) Sm{\it A}--Sm{\it C} phase transition obtained using the
parameters $w_1=-0.05,\ w_2=-1,\ w_3=-0.9$, $w_4=-0.8$, $w_5=0$
(thin lines), $w_5=-0.05$ (dashed), $w_5=-0.1$ (solid) and
$\alpha=30^{\rm o}$. Biaxial order parameters $\Gamma$ and $G_k$
together with the normalized polarization $P_s/\rho\Delta_\mu$ are
shown in (b) for the case $w_5=-0.05$ and in (c) for
$w_5=-0.1$}\label{Fig6}
\end{figure}

Now the orientational order parameters can be evaluated by numerical
minimization of the free energy at a given temperature, and the
spontaneous polarization can be calculated using the general
Eq.(\ref{Ps}) for the orientational distribution function (\ref{f1})
and the sum of Eqs.~(\ref{Umfaa}) and (\ref{UMFTc}) as the total
mean-field potential. Typical results of these calculations are
presented in Fig.~\ref{Fig6}.

Thus the present general theory enables one to calculate the
spontaneous polarization explicitly as a function of temperature
using the model of rod-like molecules with chiral distribution of
dipoles. In this case the ferroelectric ordering also occurs
according to the general mechanism described in the previous
sections. Indeed, one can readily see from Fig.~{\ref{Fig6}} that
the spontaneous polarization is approximately proportional to the
order parameter $\Gamma$ which is mainly determined by the nonchiral
part $U_0(1,2)$ of the total interaction potential. At the same
time, the spontaneous polarization is proportional to the
pseudoscalar parameter $\Delta_{\mu}\propto({\bf a}\cdot [{\bm
\mu}^+ \times {\bm \mu}^- ])$, which, of course, vanishes if the
molecules are nonchiral, for example, if both dipoles and the long
axis of the molecule presented in Fig.~{\ref{Fig5}} are in the same
plane.

\section{Conclusions}

I this paper we have used the general statistical theory and two
particular molecular models to demonstrate how the ferroelectric
ordering of polar and chiral molecules in the Smectic C* phase may
be determined by {\it nonchiral} intermolecular interactions. It has
been shown that if the molecules are characterized  by $C_{2h}$
symmetry or lower, the spontaneous polarization is given by the
general  expression ${\bf P}_s = \rho \Delta_\mu \langle{\bf a}
\times {\bf b}\rangle$ (see Eq.(\ref{PsDelta})), which is an exact
result derived without any approximations. Here the parameter
$\Delta_\mu =(\bm \mu_{\bot} \cdot ({\bf a} \times {\bf b}))$
characterizes the molecular chirality determined by the relative
orientation of the transverse molecular dipole and the molecular
plane containing the long and short molecular axes, ${\bf a}$ and
${\bf b}$ respectively. For rigid molecules of $C_{2h}$ symmetry
(see Fig.{\ref{Fig1}}), the spontaneous polarization is always
proportional to the pseudovector order parameter $\langle{\bf a}
\times {\bf b}\rangle$ which is nonzero also in the nonchiral
Smectic C phase. Thus, if specific chiral and polar intermolecular
interactions are not important, the spontaneous polarization is
essentially determined by the molecular chirality coefficient
$\Delta_\mu$ and the order parameter $\langle{\bf a} \times {\bf
b}\rangle$ emerging due to nonchiral molecular interactions in the
corresponding nonchiral Smectic C phase.

One notes that this general mechanism of ferroelectric ordering
corresponds to the one considered in the Boulder model
\cite{boulder1,boulder2}. In the Boulder model, a single molecule of
sufficiently low symmetry is ordered in the binding cite which plays
the role of the effective one-particle (mean-field) potential
considered in the present statistical theory. Similar to the
effective mean-field potential, the binding site itself has exactly
the same symmetry as the nonchiral smectic C phase (see
Fig.~{\ref{Fig2}}), i.e., possesses the mirror plane and the
two-fold symmetry axis and thus is nonchiral. One notes also that,
in the first approximation, the ordering of a molecule of the
zig-zag shape in the binding site is not determined by its possible
chirality. The corresponding nonchiral molecule with a similar
overall shape will also order in the same binding site, although no
polarization will be created in this case, of course. The ordering
of such a nonchiral molecule in the binding site is described by the
same nonchiral pseudovector order parameter ${\bf w}$ which is
considered in this paper. The same general mechanism corresponds
also to the one considered in \cite{goosens1,photinos1,photinos2}
for some particular cases. At the same time, the Boulder model as
well as the models considered in \cite{photinos1,photinos2} enable
one to account for a selection of molecular conformations which fit
the site or minimize some interaction potential. This effect is not
taken into consideration in the present paper which deals with rigid
molecules.

In this paper, the general mechanism of ferroelectric ordering has
been illustrated using two particular molecular models. The first
model is based on the electrostatic interaction between anisotropic
molecular quadrupoles. Here the molecular quadrupole tensor is
assumed to be nondiagonal in the molecular frame determined by the
molecular axes ${\bf a}, {\bf b}$ and ${\bf c}$ (i.e. one of the
primary axes of the molecular quadrupole is tilted with respect to
the long axis ${\bf a}$) . The existence of the nonzero off-diagonal
element $q_{3}$ (see Eq.(\ref{q})) determines the $C_{2h}$ symmetry
of the molecule. If this off-diagonal element is small, it is
possible to expand the orientational distribution function in powers
of $q_3$ and obtain the explicit analytical expression for the
spontaneous polarization proportional to $q_3$. This models shows
that the electrostatic quadrupole-quadrupole interaction, which is
not sensitive to molecular chirality, may be responsible for the
ferroelectric ordering of chiral molecules in the Smectic C* phase.

Finally we have considered a more realistic molecular model related
to the materials recently synthesized by Lemieux et.al.
\cite{Lemieux1999, Lemieux2001, Lemieux2002, Lemieux2005} , in which
the molecular chirality is mainly determined by the distribution of
permanent dipoles. We consider a simple model of a uniaxial rod with
two nonparallel off-center dipoles which make approximately an angle
of 90$^o$. In this simple case, the pair of permanent dipoles is
responsible for both molecular chirality, polarity and biaxiality.
The model interaction potential for such molecules is composed of
the uniaxial interaction responsible for the tilt in the smectic C
phase, and the electrostatic interaction between all dipoles. We
obtain that here the ferroelectric ordering also follows the general
mechanism described above, and the spontaneous polarization is
proportional to the pseudovector  order parameter ${\bf w}$. The
polarization is also proportional to the pseudoscalar parameter
$\Delta_\mu $ which vanishes if we set the molecule to be nonchiral
by placing the two dipoles and the long molecular axis ${\bf a}$
within the same plane. One notes that for real materials of the type
reported in \cite{Lemieux1999, Lemieux2001, Lemieux2002,
Lemieux2005}, the direct interaction between pairs of dipoles may
not be the only cause of spontaneous polarization. Such molecules
may also possess a zig-zag shape or have conformational states of
the corresponding symmetry. Then the molecule would order in the
binding site according to the Boulder model just due to steric
interactions, and this will make an additional contribution to the
spontaneous polarization. Which contribution is predominant for
particular materials can be determined by experiments involving
systematic variation of the molecular structure.

\acknowledgements The authors are grateful to R.P. Lemieux for
valuable discussions, and to EPSRC (UK) for funding.

\appendix
\renewcommand{\theequation}{A\arabic{equation}}
\section{Approximation for the electrostatic interaction between pairs of dipoles}\label{appendix}

Let us consider the electrostatic dipole-dipole interaction
(\ref{U12}) between two molecules described in
Section~\ref{section4} . In order to calculate the corresponding
part of the mean-field potential (\ref{UMFdef}), one has first to
integrate (\ref{U12}) over the intermolecular distances. This
involves the integrals
\begin{equation}\label{D}
{\bf \bar{D}}({\bf a}_1, {\bf a}_2, {\bf k})=  \int_{{\bf R} \notin
\Pi({\bf a}_1, {\bf a}_2)}{\bf D}({\bf a}_1, {\bf a}_2, {\bf R})\
d^2{\bf R}.
\end{equation}
taken over the distances $\bf R$ within the smectic plane and
accounting for the steric cut-off between the rigid cores of the
molecules with long axes ${\bf a}_1$ and ${\bf a}_2$.

The tensors (\ref{D}) have important properties: they are
invariant under the following transformations
\begin{equation}
{\bf a}_1\leftrightarrow{\bf a}_2\ \ \ \mbox{and}\ \ \ {\bf a}_1,
{\bf a}_2\leftrightarrow-{\bf a}_1, -{\bf a}_2,
\end{equation}
and are traceless and symmetric:
\begin{equation}
\delta_{ij}\bar{D}_{ij}=0,\ \ \ \bar{D}_{ij}=\bar{D}_{ji}.
\end{equation}
Furthermore, the substitutions
\begin{equation}\label{updown}
{\bf a}_1, {\bf a}_2\leftrightarrow -{\bf a}_1, {\bf a}_2\ \
\mbox{and} \ \ {\bf a}_1, {\bf a}_2\leftrightarrow {\bf a}_1, -{\bf
a}_2
\end{equation}
transform ${\bf \bar{D}}_{\uparrow\uparrow}$ into ${\bf
\bar{D}}_{\uparrow\downarrow}$ and vice versa.

We are interested in the orientational interaction which
containing the lowest possible (i.e. first) power of ${\bf
a}_{1,2}$. There are four tensorial expressions obeying the above
requirements, and thus one can approximately present the tensor
${\bf D}$ in the following form:
\begin{multline}\label{D14}
 r_0\ \bar{D}_{ij}({\bf a}_1, {\bf a}_2, {\bf
k})= \\D_1\left(a_{1i}a_{2j}+a_{2i}a_{1j}-2\ ({\bf
a}_1\cdot {\bf a}_2)\ \delta_{ij}/3\ \right)+\\
(k_ik_j-\delta_{ij}/3)[D_2\ ({\bf a}_1\cdot {\bf a}_2)+D_3\ ({\bf
a}_1\cdot {\bf k})({\bf a}_2\cdot {\bf k})]+\\
D_4[({\bf a}_1\cdot {\bf k})(a_{2i}k_j+a_{2j}k_i)+({\bf a}_2\cdot
{\bf k})(a_{1i}k_j+a_{1j}k_i)- \\4 \delta_{ij}\ ({\bf a}_1\cdot {\bf
k})({\bf a}_2\cdot {\bf k})/3].
\end{multline}
Note that we have introduced a characteristic scale parameter,
molecular breadth $r_0$ which is used to make the constants $
D_{1-4}$ dimensionless.

In order to evaluate of the constants $D_{1-4}$ it is convenient
to convolute the tensor (\ref{D14}) with another second rank
tensor. There are several possible second-rank tensors which can
be constructed from the  vectors ${\bf a}_1, {\bf a}_2$ and ${\bf
k}$. We use the simple tensor $a_{1i} a_{2j}-\delta_{ij}$, which
is orthogonal to the long axes of both molecules in the case of
ideal nematic order. It results in a convenient form of the
convolution product:
\begin{multline}\label{tildeD}
\tilde{D}({\bf a}_1, {\bf a}_2, {\bf k})= r_0\ {\bf \bar{D}:}({\bf
a}_1\ {\bf a}_2-{\bf I})=\\D_1+D_4\ [({\bf a}_1\cdot {\bf
k})^2+({\bf a}_2\cdot {\bf k})^2]+({\bf a}_1\cdot {\bf a}_2)^2\
[D_1/3-D_2/3]+\\({\bf a}_1\cdot {\bf a}_2)({\bf a}_1\cdot {\bf
k})({\bf a}_2\cdot {\bf k})[D_2-D_3/3+2D_4/3]+ \\D_3({\bf a}_1\cdot
{\bf k})^2({\bf a}_2\cdot {\bf k})^2.
\end{multline}
\begin{figure}
\centering\includegraphics[width=8cm]{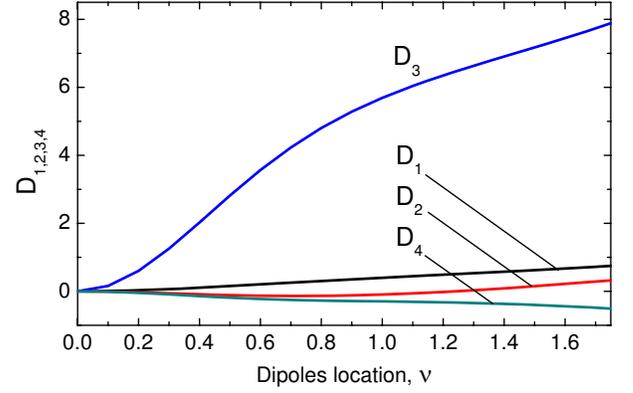}
\caption{Coefficients $D_{1-4}$ for the tensor ${\bf
D}_{\uparrow\uparrow}$ as functions of the dipole location on the
molecular long axis. The molecular axial ratio is set to be equal to
4.}\label{upup}
\end{figure}
Eq.(\ref{tildeD}) can be expressed  as a function of the molecular
tilting angles $\gamma_{1,2}$ and the difference of molecular
azimuthal angles $\phi=\varphi_2-\varphi_1$:
\begin{multline}
\tilde D (\gamma_1, \gamma_2, \phi)=d_0 + d_1
\left[P_2(\cos\gamma_1)+P_2(\cos\gamma_2)\right]+ \\d_2\
P_2(\cos\gamma_1)\ P_2(\cos\gamma_2)+ d_3\ \sin^2\gamma_1\
\sin^2\gamma_2\ \cos 2\phi+\\d_4\ \sin 2\gamma_1\ \sin 2\gamma_2 \
\cos\phi\,
\end{multline}
where the coefficients are
\begin{eqnarray}
d_0=10 D_1/9-2 D_2/27+20 D_4/27,\\
d_1=2 D_2/9+4 D_3/27+22 D_4/27,\\
d_2=2 D_1/9+2 D_2/9+8 D_3/27 +8 D_4/27,\\
d_3=D_1/6-D_2/6;\\
d_4=D_1/6+D_2/12-D_3/12+D_4/6.
\end{eqnarray}

It is possible to determine the coefficients $d_{1-4}$ by
calculating numerically the coefficients of the spherical harmonic
representation of $\tilde D$ (see e.g. \cite{uniaxial1} for
details). Then the coefficients $D_{1-4}$ can be expressed as
\begin{eqnarray}
D_1 = -d_1+5d_2/4+5d_3/3+8d_4/3,\\
D_2 = -d_1+5d_2/4-13d_3/3+8d_4/3,\\
D_3 = 9d_2/4+d_3-4d_4,\\
D_4 = 3d_1/2-3d_2/4+d_3.
\end{eqnarray}

In Figure \ref{upup}  the coefficients $D_{1-4}$ for ${\bf
D}_{\uparrow\uparrow}$ are presented. They have been calculated
for the molecules with ellipsoidal core. One can readily see that
the coefficient $D_3$  clearly dominates in $\bf D$, and thus one
may neglect other coefficients and write
\begin{equation}
\bar{D}_{\uparrow\uparrow\ ij}\approx r_0^{-1}\ D_3 \ ({\bf
a}_1\cdot {\bf k})({\bf a}_2\cdot {\bf k})(k_ik_j-\delta_{ij}/3).
\end{equation}

Obviously, the approximate form (\ref{D14}) changes sign under the
transformations (\ref{updown}), which means that ${\bf
\bar{D}}_{\uparrow\downarrow}=-{\bf \bar{D}}_{\uparrow\uparrow}$.
As a result one can write the integrated interaction energy as
\begin{multline}\label{Ubc}
\bar{U}_\mu(1,2)=\int U_\mu(1, 2)\  d^2{\bf R}=\\({\bm \mu}_1^+-{\bm
\mu}_1^-) \cdot {\bf \bar{D}}_{\uparrow\uparrow}\cdot({\bm
\mu}_2^+-{\bm \mu}_2^-)
\end{multline}

Substituting the dipole moments (\ref{mupm}) one arrives at  the
potential which depends on both long and short molecular axes. The
uniaxial part which depends only on the long axes provides a small
correction to initial uniaxial interaction potential, and we omit
it here. Secondly, there is a part which depends on a coupling of
the short axes of the two molecules:
\begin{multline}\label{Ubb}
\bar{U}^{\rm bb}(1, 2)=\frac{2\mu^2}{r_0} D_3 \sin^2\alpha ({\bf
a}_1\cdot {\bf k})({\bf a}_2\cdot {\bf k})\times
\\\left[({\bf b}_1\cdot {\bf k})({\bf b}_2\cdot {\bf
k})-\frac{1}{3}({\bf b}_1\cdot {\bf b}_2)\right].
\end{multline}

Finally there exists the most important contribution which contains
terms describing the coupling of the short axis of one molecule with
the long axis of the other. This part is responsible for the
induction of biaxial ordering by the tilt of long axes  which is of
primary importance for the description of ferroelectricity in the
smectic C* phase. This part of the potential can be written in the
form
\begin{multline}\label{Uab}
\bar{U}^{\rm ab}(1, 2)= \frac{\sqrt{2}\mu^2}{r_0} D_3\ \sin2\alpha
({\bf a}_1\cdot {\bf k})({\bf a}_2\cdot {\bf k})[({\bf b}_1\cdot
{\bf k})({\bf a}_2\cdot {\bf k})+\\({\bf a}_1\cdot {\bf k})({\bf
b}_2\cdot {\bf k})-\frac{1}{3}({\bf b}_1\cdot {\bf
a}_2)-\frac{1}{3}({\bf a}_1\cdot {\bf b}_2)],
\end{multline}

The corresponding contribution to the mean-field potential is
obtained after averaging of (\ref{Ubb}) and (\ref{Uab}) over all
orientations of the molecule '2', $U_{\rm MF}(1)=\rho\ \langle
\bar{U}(1,2)\rangle_2$, which yields
\begin{multline}\label{Umf}
U_{\rm MF}^{\rm bb}(1)=\frac{2\mu^2}{r_0} \rho\  D_3\  \sin^2\alpha\
\times
\\ \left[\frac{2}{3}G_k({\bf a}_1\cdot {\bf k})({\bf b}_1\cdot {\bf
k})-\frac{1}{3}\Gamma({\bf a}_1\cdot {\bf k})({\bf b}_1\cdot {\bf
t})\right],
\end{multline}
and
\begin{multline}\label{Umfae}
U_{\rm MF}^{\rm ab}(1)=\frac{2\sqrt{2}\mu^2}{9r_0}\ \rho\  D_3\
\sin2\alpha\ \times\\
[(2S+1)({\bf a}_1\cdot {\bf k})({\bf b}_1\cdot {\bf
k})+G_k(2P_2(\cos\gamma)+1)- \\\frac{3}{4}V ({\bf a}_1\cdot {\bf
k})({\bf b}_1\cdot {\bf
t})-\frac{3}{4}\Gamma\sin2\gamma\cos\varphi],
\end{multline}
where the order parameters $G_k=\langle({\bf a}\cdot {\bf k})({\bf
b}\cdot {\bf k})\rangle$ and $\Gamma=\langle({\bf a}\cdot {\bf
k})({\bf b}\cdot {\bf t})\rangle$ have been introduced.
Apparently, the terms containing $V\ ({\bf a}_1\cdot {\bf k})({\bf
b}_1\cdot {\bf t})$ and $\Gamma\sin2\gamma\cos\varphi$ induce the
biaxial order parameter $\Gamma$ below the tilting transition.

The order parameter $G_k$ is of minor importance, since it is
nonzero already in the Sm{\it A} phase. We have found that it is
normally of the order of 0.1 and does not affect the transition
significantly. The biaxial potential is significantly simplified
if one neglects all  terms containing $G_k$ and $({\bf a}\cdot
{\bf k})({\bf b}\cdot {\bf k})$. The corresponding simplified
mean-field potential can then be written as
\begin{multline}\label{UMFTC}
U_{\rm MF}^{\rm \Gamma}=w_5[\cot\alpha\ V({\bf a}\cdot {\bf k})({\bf
b}\cdot {\bf t})+\\\cot\alpha\ \Gamma\ \sin2\gamma\cos\varphi\
+\sqrt{2}\Gamma\ ({\bf a}\cdot {\bf k})({\bf b}\cdot {\bf t})],
\end{multline}
where the parameter $w_5=-\sqrt{2}/3\ \rho \mu^2 r_0^{-1}  D_3\
\sin^2\alpha$ is negative and can be easily estimated for given
values of molecular breadth $r_0$, 2D molecular number density
$\rho$, dipole strength $\mu$, and dipole orientation angle
$\alpha$. The reasonable values of the constant $D_3$ are between 4
and 8 (see Fig.~\ref{upup}).


\begin{thebibliography}{99}

\bibitem{Meyer1}
R.B.Meyer, L.Liebert, L.Strzelecki and P.Keller, J.Phys. (France)
lett., {\bf 36}, 69 (1975).

\bibitem{FrankJan} J.P.F. Lagerwall and F. Giesselmann,
{ChemPhysChem} {\bf 7}, 20 (2006)

\bibitem{bananareview}
G. Pelz, S. Diele, and W. Weissflog, { Adv. Matter. (Weinheim,
Ger.)} {\bf 11}, 707 (1999).

\bibitem{Helen}
S. Jaradat et al., J. Mater. Chem., 16, 3753-3761 (2006).

\bibitem{Jan}
J.P.F.Lagerwal, G.Heppke, F.Giesselmann, Eur.Phys.J. E, 18, 113
(2005)

\bibitem{osipov84}
M. A. Osipov. { Ferroelectrics} {\bf 59}, 305 (1984).

\bibitem{zeks}
B.Katnjak-Urbanc and B.Zeks, Liq.Cryst., 18, 483 (1995).

\bibitem{emelos} A.V. Emelyanenko and M.A. Osipov,
{\it Phys. Rev. E}, {\bf 68}, 051703 (2003)

\bibitem{cepic}
M. Cepic and B. Zeks, Phys. Rev. Lett. {\bf 87}, 085501 (2001).

\bibitem{ossteg}
M. A. Osipov, H. Stegemeer and A. Sprick. { Phys. Rev. E } {\bf 54},
6387 (1996).

\bibitem{MCLCreview} L. A. Beresnev, L. M. Blinov, M. A. Osipov, S. A. Pikin.
  Ferroelectric Liquid Crystals
 {\it Mol. Cryst. Liq. Cryst.}
Special Topics XXIX {\bf 158 A}, 1 - 150, (1988).

\bibitem{ospikrev}
S. A. Pikin and M. A. Osipov. Theory of Ferroelectricity in Liquid
Crystals. In {\it "Ferroelectric Liquid Crystals. Principles,
Properties and Applications". Ferroelectricity and Related
Phenomena. Vol.7}, Gordon and Breach Sci. Publ., 1992


\bibitem{Vertogen}
B.W. van der Meer and G. Vertogen, in 'Molecular Physics of Liquid
Crystals', ed. by G.R.Luckhurst and G.W.Gray (Acad. press, London
1979)

\bibitem{molreview}
M.A.Osipov, in 'Liquid Crystalline and Mesomorphic Polymers'
(Springer-Verlag, Berlin, 1993).

\bibitem{osemelchol}
A. V. Emelyanenko, M. A. Osipov and D. A. Dunmur {Phys. Rev. E,}
{\bf 62}, 2340 (2000)

\bibitem{boulder1}
M.A.Glaser, V.V.Ginzburg, N.A.Clark, E.Garcia, D.M.Walba and
R.Malzbender, Mol.Phys.Rep., {\bf 10}, 26 (1995).

\bibitem{boulder2}
D.M.Walba, 'Advances in the Synthesis and Reactivity of solids',
vol.1, 173, JAI Press Ltd. 1991.

\bibitem{photinos1}
D.J.Photinos and E.T.Samulski, Science, 270, 783 (1995).

\bibitem{photinos2}
A.F.Terzis, D.J.Photinos and E.T.Samulski, J.Chem.Phys., 107, 4061
(1997).

\bibitem{goosens1}
W.J.A.Goosens, Phys.Rev.A, 40, 4019 (1989).

\bibitem{goosens2}
W.J.A.Goosens, Ferroelectrics, 113, 51 (1991).

\bibitem{giesselmann}
F.Giesselmann, 'Smectic A-C phase Transitions in Liquid Crystals'
(Shaker-Verlag, Aachen, 1997).

\bibitem{Lemieux1999}
D. Vizitiu, C. Lazar, B.J. Halden, and R.P. Lemieux, J. Am. Chem.
Soc. {\bf 121}, 8229 (1999).

\bibitem{Lemieux2001}
D. Vizitiu, C. Lazar, J.P. Radke, C.S. Hartley, M.A. Glaser, and
R.P. Lemieux, Chem. Mater. {\bf 13}, 1692 (2001).

\bibitem{Lemieux2002}
C. Lazar, K. Yang, M.A. Glaser, M.D. Wand, and R.P. Lemieux,  J.
Mater. Chem. {\bf 12}, 586 (2002).

\bibitem{Lemieux2005}
C.J. Boulton, J.G. Finden, E. Yuh, J.J. Sutherland, M.D. Wand, G.
Wu, and R.P. Lemieux, J. Am. Chem. Soc. {\bf 127}, 13656 (2005).

\bibitem{meyer}
R.B.Meyer, Mol.Cryst.Liq.ryst., 40, 33 (1977).

\bibitem{wulf}
A. Wulf, {\it Phys. Rev. A}, {\bf 11}, 365 (1975).


\bibitem{Kocot}  M. D. Ossowska-Chrusciel, R. Korlacki, A. Kocot, R. Wrzalik, J.
Chrusciel, S. Zalewski,  Phys.Rev.E., {\bf 70}, 041705 (2004).


\bibitem{straley} J.P.Straley, Mol.Cryst.Liq.Cryst., {\bf 24}, 7 (1973),

\bibitem{gelbart} W. M. Gelbart and B.Barboi, Acc.Chem.Res., {\bf 13}, 290 (1980).

\bibitem{book}
M. A. Osipov. In {\it Handbook of Liquid Crystals. Vol.1}, 2nd
edition, edited by D.Demus, J.Goodby, G.W.Gray, H.-W. Spies and
V.Vill, Wiley-VCH, Weinheim,1998.

\bibitem{uniaxial1}
M.V. Gorkunov, F. Giesselmann, J.P.F. Lagerwall, T.J. Sluckin, and
M.A. Osipov, { Phys.Rev. E} {\bf 75}, 060701(R) (2007).

\bibitem{uniaxial2}
M.V. Gorkunov, F. Giesselmann, J.P.F. Lagerwall, and M.A. Osipov,
{\it Phys.Rev. E} in press (2007).


\end{thebibliography}
\end{document}